%% file: Main.tex
\documentclass[aip,reprint]{revtex4-1}

\usepackage{graphicx}
\usepackage{amsmath,amssymb}
\usepackage{mathptmx}
\usepackage{mathrsfs}
\DeclareMathAlphabet{\mathcal}{OMS}{cmsy}{m}{n}
\usepackage{dcolumn}
\usepackage{bm}
\usepackage[utf8]{inputenc}
\usepackage[T1]{fontenc}
\usepackage{physics}
\usepackage{etoolbox}

\usepackage{textcomp}
\usepackage[svgnames]{xcolor}
\usepackage[pdftex]{hyperref}
\hypersetup{
     colorlinks = true,
     linkcolor=DarkBlue,
     urlcolor=DarkBlue,
     citecolor=DarkBlue,
}

\newcommand{\etal}{\textit{et al.}}
\newcommand{\eff}{_{\text{eff}}}
\newcommand{\zpf}{_{\text{zpf}}}
\newcommand{\ts}[1]{_{\text{#1}}}
\newcommand{\me}{\mathrm{e}}
\newcommand{\chipm}{\chi_{\mbox{\scriptsize$\pm$}}}

\newcommand{\dif}{\mathrm{d}}
\newcommand{\beq}{\begin{equation}}
\newcommand{\eeq}{
\end{equation}}

\newcommand{\ree}[1]{\mathbb{R}\left\{#1\right\}}
\newcommand{\imm}[1]{\mathbb{I}\left\{#1\right\}}

\draft 

\begin{document}

\preprint{a}

\title{Three-Tone Coherent Microwave Electromechanical Measurement of a Superfluid Helmholtz Resonator} 

\author{S. Spence}
\email{stspence@ualberta.ca}
\affiliation{Department of Physics, University of Alberta, Edmonton, Alberta T6G 2E9, Canada}

\author{E. Varga}
\affiliation{Faculty of Mathematics and Physics, Charles University, Ke Karlovu 3, 121 16 Prague, Czech Republic}

\author{C. A. Potts}
\affiliation{Kavli Institute of NanoScience, Delft University of Technology, PO Box 5046, 2600 GA Delft, Netherlands}

\author{J. P. Davis}
\email{jdavis@ualberta.ca}
\affiliation{Department of Physics, University of Alberta, Edmonton, Alberta T6G 2E9, Canada}

\date{\today}

\begin{abstract}
We demonstrate electromechanical coupling between a superfluid mechanical mode and a microwave mode formed by a patterned microfluidic chip and a 3D cavity. The electric field of the chip-cavity microwave resonator can be used to both drive and detect the motion of a pure superflow Helmholtz mode, which is dictated by geometric confinement. The coupling is characterized using a coherent measurement technique developed for measuring weak couplings deep in the sideband unresolved regime. The technique is based on two-probe optomechanically induced transparency/amplification using amplitude modulation. Instead of measuring two probe tones separately, they are interfered to retain only a signal coherent with the mechanical motion. With this method, we measure a vacuum electromechanical coupling strength of $g_0 = 2\pi \times 23.3 \; \mathrm{\mu}$Hz, three orders of magnitude larger than previous superfluid electromechanical experiments. 
\end{abstract}


\maketitle 


Cavity optomechanics, \cite{aspelmeyer2014cavity} the coupling of optical and mechanical resonances, is an extremely sensitive scheme for detecting and controlling mechanical motion. Popular optomechanical architectures include vibrating membranes, \cite{yuan2015large,noguchi2016ground,pearson2020radio,pate2020casimir} superconducting drums,\cite{teufel2011sideband,kotler2021direct,cattiaux2021macroscopic} phononic crystal cavities,\cite{arrangoiz2018coupling,serra2021silicon} and magnetic spheres.\cite{zhang2016cavity,potts2021dynamical} While decidedly more exotic, superfluid helium has been explored as a promising mechanical element for cavity optomechanical and electromechanical experiments -- in part due to its large bandgap,\cite{glaberson1975impurity,kandula2010extreme} low dielectric loss,\cite{hartung2006rf} and ultra-low acoustic loss at millikelvin temperatures.\cite{de2014superfluid,souris2017ultralow}
Superfluid helium mechanical resonators have been used to develop gravitational wave detectors,\cite{singh2017detecting,vadakkumbatt2021prototype} and have even been suggested for generating a mechanical qubit \cite{sfendla2021extreme} and for detecting dark matter,\cite{manley2020searching} while superfluid optomechanical resonators are allowing new studies of superfluid properties, such as novel explorations of vortex dynamics.\cite{sachkou2019coherent}  Microfabricated superfluid mechanical resonators --- to date lacking the optomechanical component --- have also proven to be particularly useful in the study of confined helium, where confinement can dramatically change the physics of the system.\cite{levitin2013phase,perron2019review} Previously, our group has developed microfluidic Helmholtz resonators for this purpose,\cite{duh2012microfluidic,rojas2015superfluid,souris2017ultralow} which have been used in $^4$He to discover bi-stable turbulence in 2D superflow \cite{varga2020observation} and reveal surface-dominated finite-size effects at the nanoscale.\cite{varga2022surface} These devices have also been used to search for pair density wave states in superfluid $^3$He.\cite{shook2020stabilized} Here, we combine the concepts of microwave cavity electromechanics with microfluidic confinement of superfluid helium-4 within a microfabricated Helmholtz resonator. 

\begin{figure}
\includegraphics[width=0.48\textwidth]{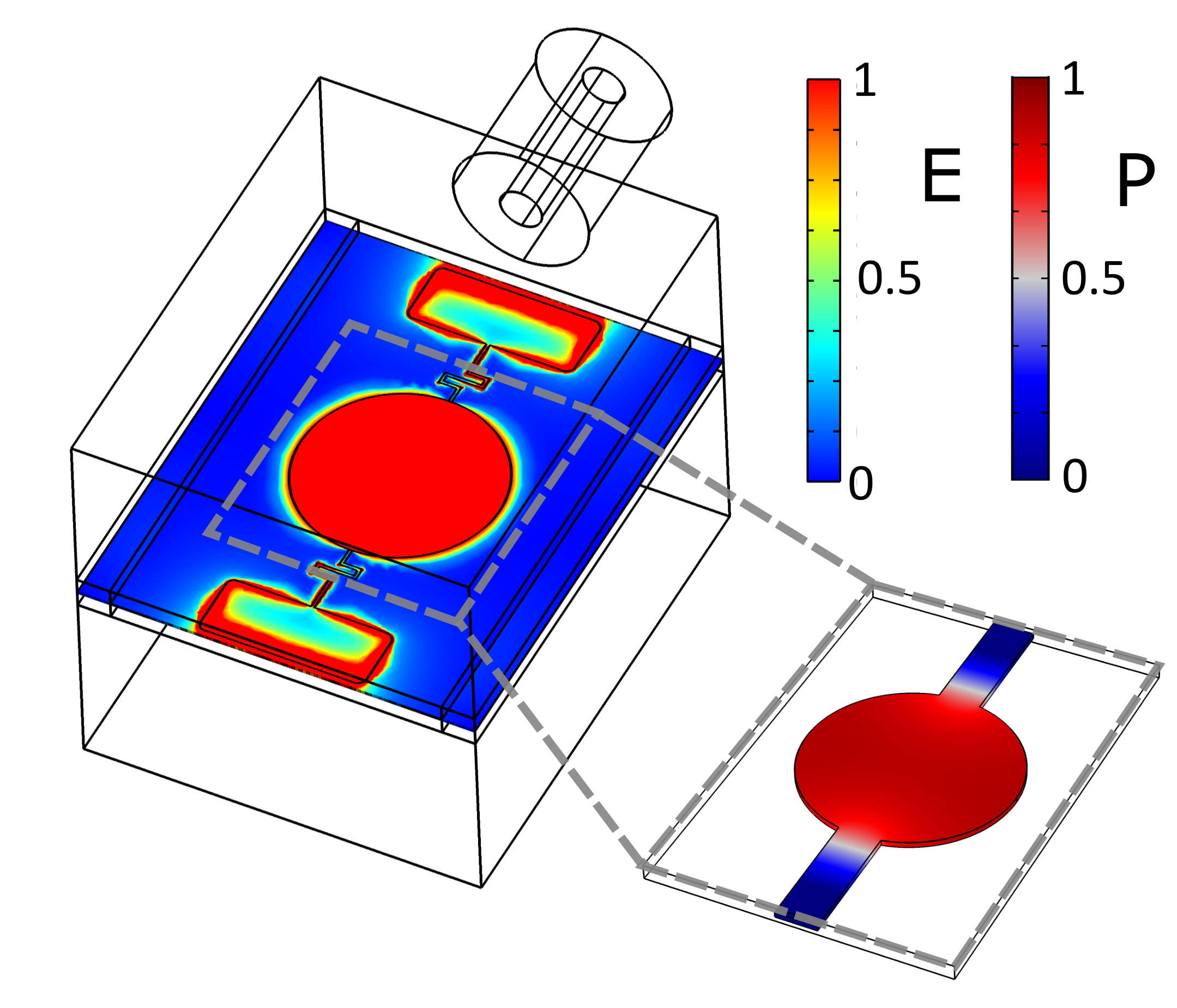}
\caption{\label{fig:one} The chip-cavity system. The outline shows the 3D microwave cavity and pin coupler, while the relative electric field magnitude of the fundamental chip-cavity microwave mode, concentrated between the central basin capacitive plates and around the antenna, is shown according to the left color bar. Insert: the relative pressure of the superfluid Helmholtz mode within the fluid geometry of the chip. Pressure is concentrated in the central basin, with relative amplitude according to the right color bar.}
\end{figure}

In the present work, a helium volume is enclosed between two nanofabricated quartz substrates, defining a fourth sound\cite{tilley2019superfluidity} (pure superflow) Helmholtz mode inside the microfluidic chip.\cite{duh2012microfluidic,rojas2015superfluid,souris2017ultralow} The pressure field of this mode is shown in the insert of Fig.~\ref{fig:one}. Previously, Varga \etal\cite{varga2021electromechanical} showed the value in electromechanical drive and detection of the Helmholtz mode using kHz-frequency carrier signals in a cavity-less system.  Here, we bring the readout into the GHz microwave regime by integrating the chip into a hermetically-sealed 3D microwave cavity filled with superfluid $^4$He, with on-chip superconducting aluminum electrodes concentrating the cavity's electric field into the Helmholtz basin (main image Fig.~\ref{fig:one}), similar to work with SiN membrane chips in 3D microwave cavities.\cite{yuan2015large,noguchi2016ground} A three-tone coherent drive and measurement technique was developed to observe small mechanical signals and calibrate the optomechanical coupling rate, $g_0$, using a strong pump microwave tone which is amplitude modulated to give two weak probe tone sidebands. The chip-cavity system's effect on the two sidebands is analogous to optomechanically induced transparency and amplification (OMIT/A).\cite{weis2010optomechanically} Demodulating the signal to destructively interfere the two probe tones recovers only the coherent microwave signal containing information about the mechanical motion. This provides a powerful measurement technique in the sideband unresolved regime, especially useful for detecting mechanics in systems with weak opto/electro-mechanical couplings. We use this technique to characterize the chip-cavity electromechanical coupling, measuring a vacuum electromechanical coupling of $g_0 = 2\pi \times 23.3 \; \mathrm{\mu}$Hz.


\begin{figure*}
\includegraphics[width=0.9\textwidth]{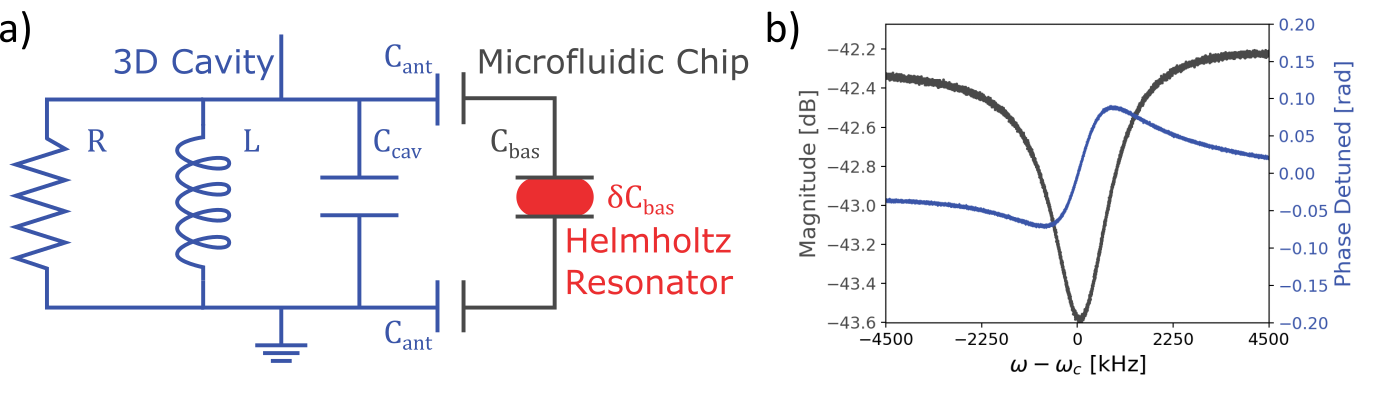}
\caption{\label{fig:two} Left: an RLC representation of the chip-cavity system, showing the 3D microwave cavity in blue, the microfluidic chip in black, and the superfluid Helmholtz mode in red. Right: A typical reflection measurement (via a directional coupler) of the cavity at 600 mK when filled with superfluid $^4$He. The magnitude is shown in grey, while the detuned phase is shown in blue. Frequency has been detuned around the microwave cavity frequency $\omega_c = 2\pi \times 2.28$ GHz.}
\end{figure*}

As mentioned above, the experimental system consists of a superfluid Helmholtz resonator, similar to Refs.~\onlinecite{souris2017ultralow,varga2020observation,varga2021electromechanical}, but now incorporated into a hermetic 3D microwave cavity. The etched Helmholz geometry is a wide flat circular basin, 7 mm in diameter and 1.01 $\mathrm{\mu}$m deep, connected by two 1.6 mm wide 1.6 mm long channels to the surrounding helium bath. Electrodes of 50 nm thick aluminum are deposited into the etched regions:  a 6 mm diameter circle at the center of the basin connects to a meandering 100 $\mathrm{\mu}$m-wide lead (in one channel) that terminates in a 5 mm $\times$ 2 mm square antenna. After fabrication, the wafer is diced into identical top and bottom chips, which are room-temperature direct wafer bonded \cite{tong1994low,plach2013mechanisms} such that the basins align and the antenna are on opposite sides of the basin (as shown in Fig.~\ref{fig:one}). The combined basins form a single volume with aluminum top and bottom electrodes, turning the combined basin into a parallel plate capacitor, with an antenna connected to each plate -- providing capacitive coupling to the 3D cavity mode (more detailed fabrication notes can be found in Ref.~\onlinecite{souris2017ultralow}).

The Helmholtz mode is an acoustic resonance of the superfluid moving back and forth in the channel, driven by pressure fluctuations in the central basin (shown in the insert of Fig.~\ref{fig:one}). The mode can be described as a mass-spring system,\cite{varga2020observation} with effective mass $m_{\text{eff}} = 2 w l D \rho_{\text{s}}$; where $w$, $l$ and $D$ are the channel width, length, and depth respectively; and $\rho_{\text{s}}$ is the superfluid density. The resulting mode frequency is then:\cite{varga2021electromechanical}

\begin{equation}
\label{eq:Omegam}
    \Omega\ts m = \sqrt{\frac{k\eff}{m\eff}} =  \sqrt{\frac{2 w D}{l\rho} \frac{\rho_{\text{s}}}{\rho} \frac{k_{\text{p}}}{4 A^2 (1+\Sigma)}} ,
\end{equation}
with $k\eff$ the effective spring constant, $k_{\text{p}}$ the effective stiffness of the mean deflection of the basin walls, $A = \pi R^2$ the area of the basin, and $\Sigma = \chi_{\text{B}} D k_{\text{p}}/(4A)$, where $\chi_{\text{B}}$ is the bulk compressibility of helium. Here, the effective spring constant of the mode is significantly softened by the flexing of the substrate when compared to the compressibility-only case. The flexing of the substrate changes the distance between the capacitive plates above and below the basin, the origin of the electromechanical coupling for this system, which is slightly suppressed by electrostriction due to the compression of the helium.\cite{de2016optomechanics,spence2022superfluid} Considering both of these factors, the change in basin capacitance $C\ts{bas}$ can be written in terms of the displacement of the fluid in the channels $y$, as:\cite{varga2021electromechanical}

\begin{equation}
\label{eq:dcdy}
    \frac{\dif C\ts{bas}}{\dif y} = \bar{C}\ts{bas} \frac{2 w \rho_{\text{s}}}{A \rho} \left( 1 - \frac{\varepsilon - 1}{\varepsilon}\Sigma \right) \frac{1}{1 + \Sigma} ,
\end{equation}
where $\bar{C}\ts{bas}$ is the undeformed basin capacitance and $\varepsilon$ is the dielectric constant of liquid helium. 

The Helmholtz chip is placed at the center of a rectangular 3D microwave cavity,\cite{pozar2011microwave} and the 3D cavity mode is capacitively coupled to the basin capacitor via the on-chip antenna,\cite{schuster2007circuit} effectively concentrating the electric field of the fundamental chip-cavity mode into the basin, as shown in Fig.~\ref{fig:one}. The lumped element RLC representation of the chip-cavity system is shown in Fig.~\ref{fig:two} (a), and (b) shows a typical $S_{21}$ single port bi-directional (reflection) measurement of the microwave resonance at 600 mK with the cavity filled with superfluid helium-4 close to saturated vapor pressure (SVP). Microwave circuit schematics can be found in the Supplementary Information (SI). From this measurement, the resonant frequency is found to be $\omega\ts c = 2\pi \times 2.28$ GHz, with total cavity loss rate $\kappa = 2\pi \times 1.76$ MHz, and external coupling rate $\kappa_{\text{in}} = 2\pi \times 128$ kHz, meaning the system is significantly undercoupled. \cite{rieger2022fano}

The vacuum electromechanical coupling strength $g_0 = G x\zpf$, with $G = - \partial \omega\ts c / \partial x$, can be considered as the shift in cavity resonance due to the zero point motion of the mechanical oscillator. For the Helmholtz chip-cavity system, this can be written as:

\begin{equation}
    g_0 = - \delta y\zpf \frac{\partial \omega\ts c}{\partial y} =  - \delta y\zpf \frac{\partial \omega\ts c}{\partial C} \frac{\partial C}{\partial y}, 
\end{equation}
where $y\zpf$ is the zero point motion of the superfluid in the channels, and $C$ is the total capacitance of the chip-cavity microwave mode given by $\omega\ts c = 1/\sqrt{LC}$. Using $\delta y\zpf = \sqrt{\hbar / 2 m\eff \Omega\ts m}$, Eq.~\ref{eq:dcdy}, and that the temperature is cold enough for $\rho\ts s \approx \rho$, $g_0$ can be expressed for this electromechanical coupling scheme as:

\begin{equation}
\label{eq:g0theory}
    g_0 \simeq \alpha \omega\ts{c} \sqrt{\frac{\hbar}{2 m\eff \Omega\ts m}} \frac{w}{A} \left( 1 - \frac{\varepsilon - 1}{\varepsilon}\Sigma \right) \frac{1}{1 + \Sigma}.
\end{equation}
Here, $\alpha$ is the proportion of the microwave mode's total electric field energy within the basin capacitor; for this work, $\alpha \approx 0.005$ is calculated via finite element method (FEM) simulations. Equation~\ref{eq:g0theory} gives a method for estimating the $g_0$ of a fabricated device. Only $\Omega\ts m$ and $k\ts p$ are unknown, yet $\Omega\ts m$ can be estimated using Eq.~\ref{eq:Omegam} while $k\ts p = 9.45 \times 10^6$ N m$^{-1}$ is measured via capacitance change with a DC bias, as in Ref.~\onlinecite{souris2017ultralow}. Using these values an estimate of $g_0 \approx 2\pi \times 1.1 \; \mu$Hz is obtained; although lower than the value found later in this work, this discrepancy is likely due to inaccuracy in $\alpha$.


We began by observing electromechanical coupling using a calibrated homodyne measurement:\cite{gorodetksy2010determination,kumar2023novel,spence2022superfluid} sending a tone at the microwave resonance's maximum gradient, demodulating the returned signal with a tone of the same frequency, measuring noise in the time domain, and considering the power spectral density of fluctuations near the mechanical frequency $\Omega\ts m$ (circuit shown in the Supplementary Information). Reflection via a directional coupler was measured as the return signal. The mechanical spectrum was found using an amplitude-sensitive scheme, giving a Helmholtz mode frequency of $\Omega\ts m = 2\pi \times 1394.7$. The homodyne phase-lock was achieved by applying a 10 kHz phase modulation to the signal on the cavity arm of the circuit and then minimizing the 10 kHz signal after down-mixing by varying the LO phase. Phase-sensitive measurement was also possible by instead sending the microwave tone on resonance and using 10 kHz amplitude modulation to phase-lock.

Despite homodyne measurement enabling detection of the mechanical spectrum, the signal is weak, with a small signal-to-noise ratio near the mechanical resonance frequency, mainly due to low electromechanical coupling strength and undercoupled microwave readout, even after more than 450 averages. Moreover, determination of $g_0$ was not possible due to uncertainty in the effective temperature of the Helmholtz mode since vibrations from the cryocooler drove the superfluid motion in the present experiment. 


\begin{figure*}
\includegraphics[width=0.95\textwidth]{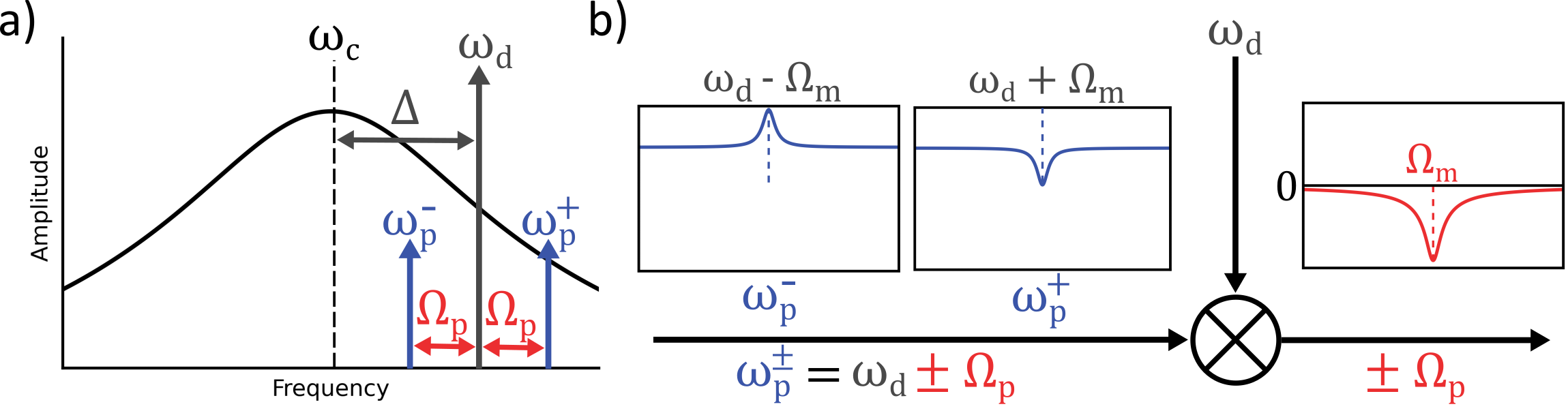}
\caption{\label{fig:three} Diagram of the three-tone coherent measurement scheme. a) Frequency space diagram of the three tones relative to the microwave resonance $\omega\ts c$. Showing the pump tone $\omega\ts d$ at detuning $\Delta$ from the cavity resonance, along with the two probe tones $\omega^\pm \ts p$ created via amplitude modulation of the pump tone with frequency (separation) $\Omega\ts p$. The cavity resonance lineshape has been exaggerated for clarity. b) Diagram of the two probe tones before (left - blue) and after (right - red) demodulation. The left side shows the response at the probe tones $\omega^\pm \ts p$ as the modulation frequency $\Omega\ts p$ is swept across the mechanical frequency $\Omega\ts m$, showing the EMIT/A response for the upper and lower sidebands respectively. The flat background is due to being deep into the sideband unresolved regime $\kappa \gg \Omega \ts m$. The right side shows the measured signal at relative phase $\phi = 0$, after demodulating the two probe tones with the pump frequency $\omega\ts d$. This down-mixes the probe signals to frequency $\pm \Omega \ts p$, where the two probe tones destructively interfere, leaving only the combined EMIT/A signal with a peak at $\Omega\ts m$.}
\end{figure*}

A coherent measurement scheme can be used to precisely determine the electromechanical coupling rate, $g_0$. The scheme in this work is based on OMIT/A:\cite{weis2010optomechanically,agarwal2010electromagnetically,safavi2011electromagnetically,kashkanova2017superfluid} an interference effect where beating between two optical tones causes coherent oscillation of the mechanics via radiation pressure when the two tones are separated in frequency by the mechanical resonance. The oscillation of the mechanics, in turn, scatters photons from the stronger driving `pump' tone to the weak `probe' tone frequency, where they then destructively (constructively) interfere with the probe tone in the case of OMIT (OMIA). In the microwave regime, the equivalent effect is often called electromechanically induced transparency/amplification (EMIT/A).\cite{zhou2013slowing} For the Helmholtz devices, mechanical motion is driven via modulation of the electrostatic force between the two capacitive basin plates, proportional to voltage squared. Typically OMIT/A is used for sideband resolved systems ($\kappa \ll \Omega\ts m$), with stronger optomechanical coupling strengths compared to this work. To overcome these limitations, a `two-probe' measurement scheme is used here. The two probe tones are produced by amplitude modulation of the pump tone, inspired by the work of Kashkanova \etal,\cite{kashkanova2017optomechanics} and adapted from optics to the microwave regime. For this work, modulation of a microwave signal allows straightforward resolution of a mHz mechanical linewidth on a GHz signal and, for kHz mechanics deep into the sideband unresolved regime, enables a unique measurement scheme where the probe tones are canceled to recover only a coherent electromechanical signal, which would not be visible otherwise.

Figure~\ref{fig:three} shows a diagram of the coherent measurement scheme in this work. In Fig.~3(a), a strong microwave pump tone with frequency $\omega\ts d$ is sent into the cavity, at some detuning $\Delta$ from the cavity resonance $\omega\ts c$, such that $\Delta = \omega\ts d - \omega\ts c$. This pump tone is amplitude modulated with frequency $\Omega\ts p$ to produce the two weak probe signals at $\omega^{\pm}\ts p = \omega\ts d \pm \Omega\ts p$. The signal incident at the cavity port can then be written as $s\ts{in} = \bar{s}\ts{in} + \delta s\ts{in}$, where the weak probes provide $\delta s\ts{in}$, which in a frame rotating at $\omega\ts d$ can be written:

\begin{equation}
    \delta s\ts{in} = s\ts p \left( \me^{-i \Omega\ts p t} + \me^{+i \Omega\ts p t} \right)/2.
\end{equation}
Here, $s\ts p/2$ is the amplitude of the probe signals. Sweeping $\Omega\ts p$ across the mechanical frequency $\Omega\ts m$, equivalent to sweeping the probe and pump separation, causes the EMIT/A effect close to $\Omega\ts m$ for the upper and lower sidebands respectively, shown in the insert of Fig.~\ref{fig:three}(a). Following the full derivation in the Supplementary Information, the intracavity field amplitude at the two probe frequencies $\omega^{\pm}\ts p$ can be written as
\begin{equation}
    A^\pm (\Omega\ts p) = \frac{\chipm \sqrt{\kappa\ts{in}}s\ts p}{2} \left( 1 \mp \frac{g (g\chi_\mp^* +g^*\chipm)}{\Gamma\ts m /2 \mp i\delta \pm i\Sigma(\pm \Omega\ts p)} \right) , 
\end{equation}
where $\Gamma\ts m$ is the mechanical linewidth, $\delta = \Omega\ts p - \Omega\ts m$ is the detuning of the pump-probe separation from the mechanical frequency, $g$ is the multiphoton coupling defined as
\begin{equation}
    g = g_0\frac{\sqrt{\kappa\ts{in}}\bar{s}\ts{in}}{-i\Delta+\kappa/2},
\end{equation}
$\chipm$ is the cavity susceptibility at $\omega^\pm\ts p$
\begin{equation}
    \chipm = \frac{1}{-i(\Delta \pm \Omega\ts p) + \kappa/2},
\end{equation}
and the electromechanical self-energy $\Sigma(\Omega\ts p)$ is defined as
\begin{equation}
    i\Sigma(\Omega\ts p) = \vert g \vert^2(\chi_+ - \chi_-^*).
\end{equation}

When measuring in a bi-directional single port scheme (reflection via a circulator / directional coupler) $-\sqrt{\kappa\ts{in}} A^\pm$ is effectively the modification of the reflected probe tones due to the cavity electromechanical system. Demodulating this reflected signal at $\omega\ts d$ and a relative phase of $\phi = 0$, will give a signal at $\Omega\ts p$ with amplitude proportional to the upper probe minus the lower probe, canceling out the probe tone backgrounds and recovering only the coherent effect, shown in Fig~\ref{fig:three}(b) and (c). Measuring this signal at $\Omega\ts p$ using a lock-in amplifier will give a signal
\begin{equation}
\label{eq:xplusiy}
    X+iY =\frac{- i K\ts T g_0^2\kappa\ts{in}^2 s\ts p }{\sqrt{2} (\Gamma\ts m /2 + i\delta)}  \frac{(\kappa^2/4 - \Delta^2)}{(\kappa^2/4 + \Delta^2)^3} \frac{P\ts{in}}{\hbar \omega\ts d} ,
\end{equation}
where $K\ts T$ is the transfer function of the circuit post reflection from the cavity and $P\ts{in}$ is the total power incident on the cavity port. A full derivation of $X+iY$ can be found in the Supplementary Information. Figure~\ref{fig:four}(b) shows a measurement of the $Y$ quadrature, fit to Eq.~\ref{eq:xplusiy}, showing an improvement of fitting accuracy and signal-to-noise near resonance, compared to the homodyne measurement of noise amplitude spectral density in Fig.~\ref{fig:four} (a). From the coherent measurement fit, the mechanical frequency and linewidth are found to be $\Omega\ts m = 2\pi \times 1394.7$ Hz and $\Gamma\ts m = 2\pi \times 0.106$ Hz. Although it may instead be possible to cancel the probe tones using an additional source, with the current method far into the sideband unresolved regime ($\kappa \gg \Omega\ts m$), the two probe tones effectively travel the same path, canceling near perfectly and accounting for any frequency, phase, or amplitude noise. Using the three-tone method, the background signal at $\Omega\ts m$ was reduced by three to four orders of magnitude, allowing a measurement of the mechanics that would not normally be possible with EMIT/A.

\begin{figure*}
\includegraphics[width=0.9\textwidth]{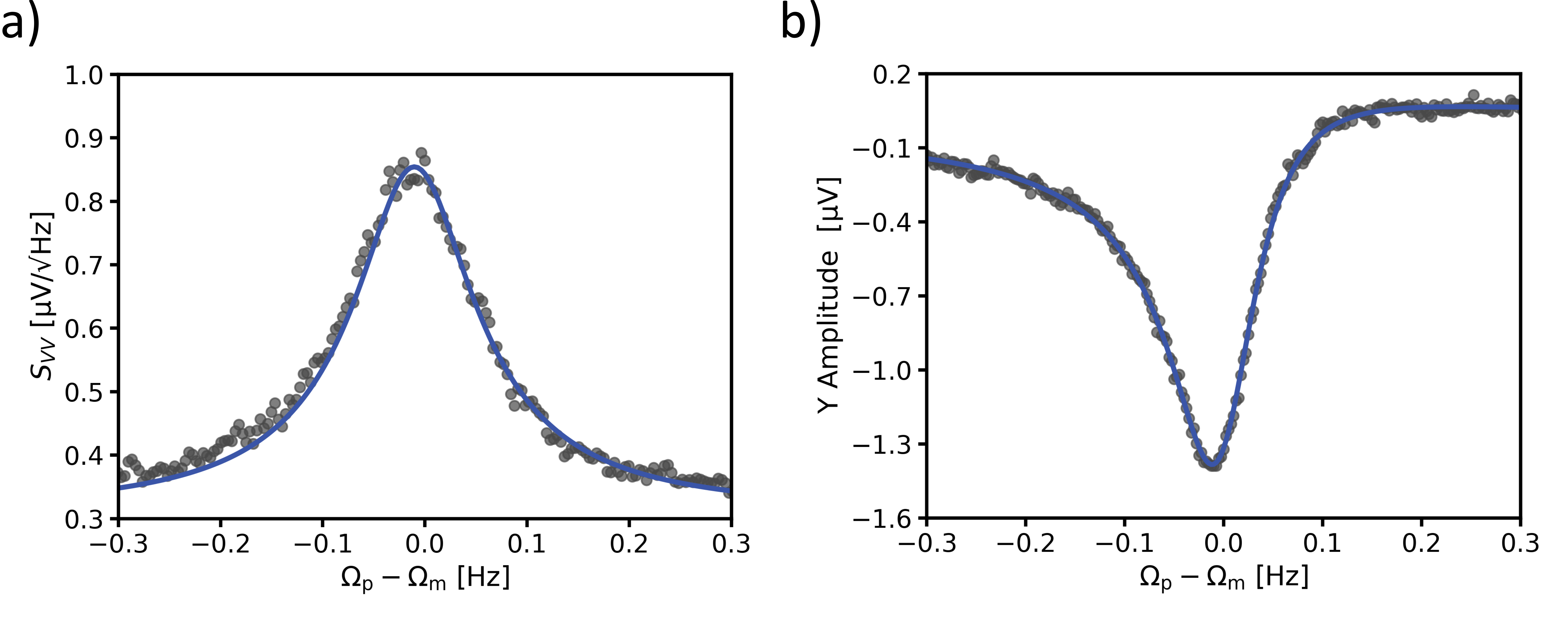}
\caption{\label{fig:four} a) Homodyne noise measurement: plotting noise amplitude spectral density of the Helmholtz mode, as driven by the environmental noise of the cryocooler. Shown is the average of 453 acquisitions. b) Coherent three-tone measurement (EMIT/A): plotting the amplitude measured at the lock-in vs. amplitude modulation frequency, detuned around the mechanical resonance $\Omega\ts m$. Shown is the average of 37 acquisitions.}
\end{figure*}

The circuit transfer function $K\ts T$ in Eq.~\ref{eq:xplusiy} can be calibrated to allow determination of $g_0$ using a measurement of $R = \vert X + iY \vert$ with $\phi = \pi/2$, to give a signal proportional to the combined probe tone amplitudes $R\ts{cal} =  K\ts T s\ts p / 2 \sqrt{2}$ (further details in the Supplementary Information). Using this calibration, $g_0$ can be written as
\begin{equation}
    g_0 = \frac{\sqrt{\hbar\omega\ts d}}{2 \kappa\ts{in} \sqrt{P\ts{in}}} \sqrt{\frac{\Gamma_m(\kappa^2/4 + \Delta^2)^3}{(\kappa^2/4 - \Delta^2)} \frac{R_{\delta =0}}{R_{cal}}}.
\end{equation}
Here, $R_{\delta=0}$ is the measured signal amplitude at $\Omega\ts p = \Omega\ts m$ when the probe tones cancel, either directly from the peak signal or from fitting to Eq.~\ref{eq:xplusiy}. Using this methodology, the vacuum electromechanical coupling strength was calculated as $g_0 = 2\pi \times 23.3^{+6.0}_{-4.8}\; \mu$Hz, with determination of $P\ts{in}$ providing the dominant error. This $g_0$ is in the weak coupling limit of electromechanics,\cite{aspelmeyer2014cavity} but a three order of magnitude increase when compared to a bulk superfluid-microwave coupling.\cite{de2017ultra} While the corresponding single-photon cooperativity $C_0= 1.2 \times 10^{-14}$ is one of the lowest recorded,\cite{aspelmeyer2014cavity} and will need to be improved in subsequent experiments, this demonstrates the sensitivity of the three-tone coherent measurement scheme. 


In conclusion, the first superfluid microwave electromechanical coupling was measured for a micromechanical system, using a microfluidic Helmholtz resonator inside a 3D microwave cavity. The mechanical motion was detected using a homodyne noise measurement scheme; however, thermomechanical calibration was not possible due to strong cryocooler vibrations. To measure an electromechanical system with weak coupling, deep in the sideband unresolved regime, a coherent measurement scheme was developed, inspired by OMIT\cite{weis2010optomechanically} and specifically the work of Kashkanova \etal\cite{kashkanova2017optomechanics} The scheme uses amplitude modulation of a strong pump tone to create probe sidebands, which experience coherent EMIT/A. The two probe tones are then canceled to recover only signal coherent with the mechanics, allowing an electromechanical calibration of the coupling strength $g_0$.

The coherent measurement scheme provides a powerful tool for measuring electro/optomechanical systems deep into the sideband unresolved regime. Moreover, integrating a superfluid Helmholtz device into a microwave cavity provides a new platform for superfluid microwave electromechanics. With three orders of magnitude improvement to $g_0$ over previous superfluid electromechanical systems,\cite{de2016optomechanics} the microwave Helmholtz design shows promise as a sensitive method for measuring the properties of superfluid helium. The sensitivity of this proof of concept can be improved by orders of magnitude in many areas: confinement could decrease $\Gamma\ts m$ while increasing $g_0$ (possibly with a phononic crystal\cite{spence2021superfluid,sfendla2021extreme}), larger antenna or galvanic bonding would increase $\alpha$,\cite{noguchi2016ground} low mK dilution temperatures would decrease superfluid and microwave dissipation, a superconducting microwave cavity could also decrease dissipation, and a stronger coupling into the cavity ($\kappa\ts{in} \approx \kappa/2$) would greatly improve signal strength for any given cavity field.  The improved system may be useful for studying non-equilibrium thermodynamics,\cite{awschalom1984observation} or even vector dark matter.\cite{manley2021searching}

The authors acknowledge that the land on which this work was performed is in Treaty Six Territory, the traditional territories of many First Nations, Métis, and Inuit in Alberta. They acknowledge support from the University of Alberta; the Natural Sciences and Engineering Research Council, Canada (Grant Nos. RGPIN-2022-03078, and CREATE-495446-17); the NSERC Alberta Innovates Advance program (ALLRP-568609-21); and the Government of Canada through the NRC Quantum Sensors Program.

\section*{Author Delclarations}
\subsection*{Conflict of Interest}
The authors have no conflicts to disclose.

\subsection*{Author Contributions}
\textbf{Sebastian Spence:} Conceptualization (supporting), Formal Analysis (lead), Methodology (equal), Software (lead), Visualization (lead), Writing (lead). \textbf{Emil Varga:} Conceptualization (lead), Formal Analysis (supporting), Methodology (equal), Supervision (supporting), Writing (supporting). \textbf{Clinton Potts:} Formal Analysis (supporting), Software (supporting), Writing (supporting). \textbf{John Davis:} Conceptualization (supporting), Funding Acquisition (lead), Supervision (lead), Writing (supporting).

\section*{Data Availability}
The data that support the findings of this study are available from the corresponding author upon reasonable request.

\bibliography{bibliography}

\onecolumngrid
\appendix

\input{SI}

\end{document}

%% file: SI.tex
\def\subinrm#1{\sb{\rm#1}}
{\catcode`\_=13 \global\let_=\subinrm}
\mathcode`_="8000
\def\upsubscripts{\catcode`\_=12 } \def\normalsubscripts{\catcode`\_=8 }
\upsubscripts

\newpage
\section*{Supplementary Information}

\section{Three-Tone Coherent Measurement.}
\label{sec:three tone}
\begin{figure}[h!]
  \centering
  \includegraphics[width=0.8\textwidth]{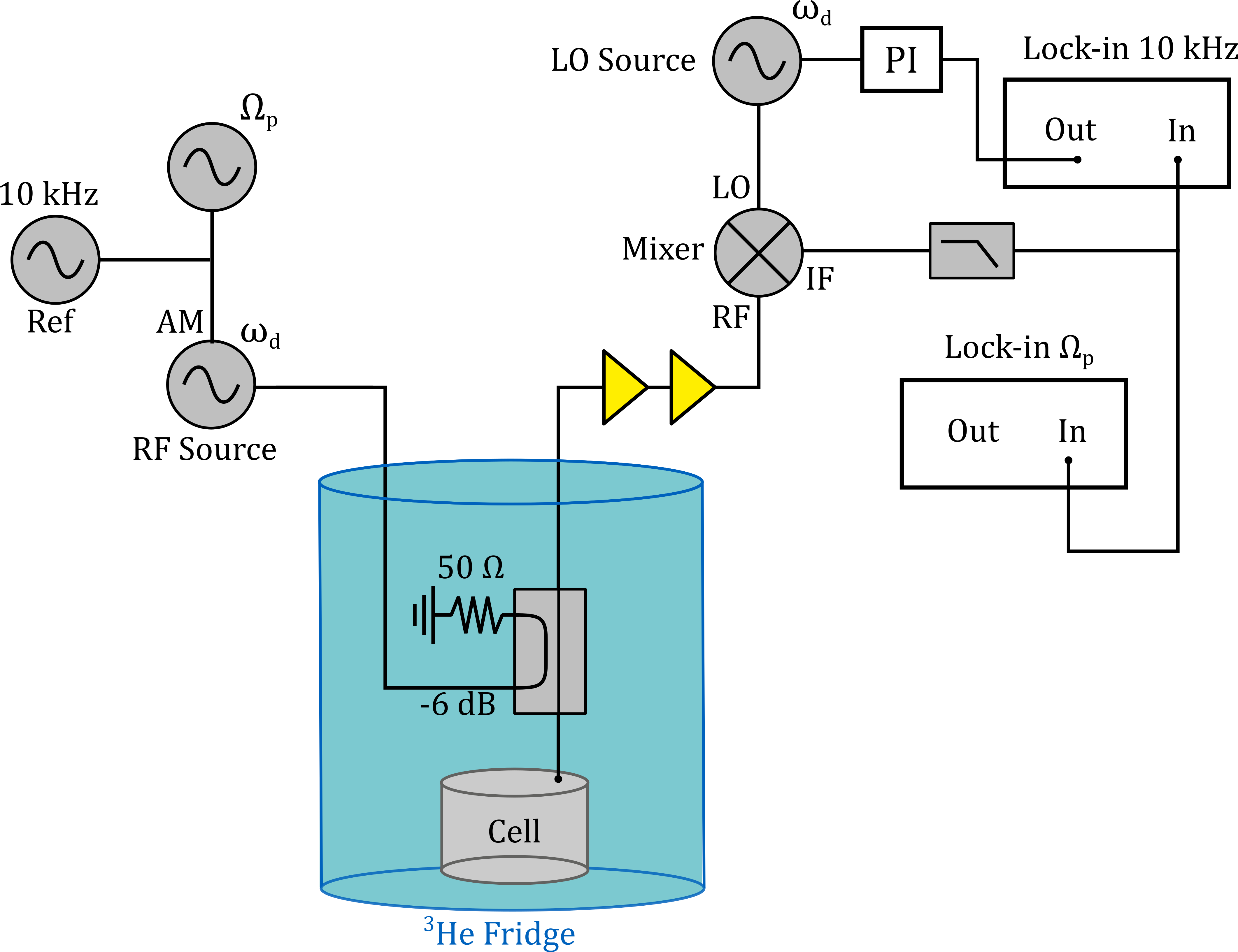}
  \caption{Three-tone coherent measurement circuit. The `RF Source' microwave signal at pump frequency $\omega_d$ is amplitude modulated with frequency $\Omega_p$ and the `10 kHz' reference signal, which is used for phase-locking. The RF signal is sent to the experimental `Cell' (cavity-Helmholtz) within the $^3$He cryostat (at 600 mK) via a directional coupler. The reflected microwave signal passes back through the directional coupler, is amplified by two room temperature microwave amplifiers, and is mixed with a signal of frequency $\omega_d$ from the `LO Source'. The mixed signal is low-pass filtered to remove the $2\omega_d$ components and then measured on two lock-in amplifiers. The lock-in set to $\Omega_p$ (lower on the diagram) measures the signal coherent with the mechanics, while the 10 kHz lock-in effectively measures the relative phase $\phi_{ref} - \phi$ at the mixer. The 10 kHz lock-in measurement is taken as the input of a PI loop, which maintains the relative phase at the mixer by adjusting the LO Source output phase.}
  \label{fig:omitcircuit}
\end{figure}

This section will detail the theory of three-tone coherent measurement, inspired by the work of Kashkanova \etal~\cite{kashkanova2017optomechanics}, adapted to microwave circuits and developed to include probe tone canceling. This technique enables sensitive measurement of weak electromechanical couplings deep into the sideband-unresolved regime. First, the theory for two-probe EMIT/A is outlined using an amplitude modulation scheme, then developed to include destructive interference of the probe tones, recovering only the signal coherent with the mechanical motion. Finally, details are provided for calibrating coherent signal strength against the constructively interfered probe tones. Figure~\ref{fig:omitcircuit} shows the microwave measurement circuit for the three-tone coherent measurement. 

\subsection{Two-Probe EMIT/A}
This subsection will outline the theory for two-probe EMIT/A using an amplitude modulation scheme. The method in this section is based on Ref.~\cite{weis2010optomechanically}, modified to include two probe beams provided by amplitude modulation of the pump tone. Refs. \cite{kashkanova2017optomechanics} and \cite{kashkanova2017superfluid} have also been used to assist with these modifications. To begin, the electromechanical Hamiltonian can be written as

\beq
\hat{\mathcal{H}} = \frac{\hat{p}^2}{2m_{eff}} + \frac{1}{2}m_{eff} \Omega_m^2 \hat{x}^2 + \hbar \omega_c \left(\hat{a}^\dagger \hat{a} + \frac{1}{2} \right)
+ \hbar G \hat{a}^\dagger \hat{a}\hat{x} + i\hbar \sqrt{\kappa_{in}} \left( s_{in}(t) \hat{a}^\dagger - s_{in}^* (t) \hat{a}\right) \; .
\eeq
Here, $\hat{p}$ and $\hat{x}$ are the momentum and position operators of the mechanical mode, with resonant frequency $\Omega_m$ and effective mass $m_{eff}$; $\hat{a}$ and $\hat{a}^\dagger$ are the creation and annihilation operators of the microwave cavity mode, with resonant frequency $\omega_c$, external loss rate $\kappa_{in}$, and incident field at the cavity port $s_{in}(t)$; and finally $G = \partial \omega_c / \partial x$ is the electromechanical coupling strength.

For a strong microwave pump (drive) tone at $\omega_d$ and weak probe tones (one or more), the incident signal can be written $s_{in}(t) = \left(\bar{s}_{in} + \delta s (t)\right)\me^{-i\omega_d t}$, where $\delta s_{in}(t)$ contains the probe tones and the pump tone amplitude $\bar{s}_{in}$ is both positive and real. Then, writing the Langevin equations in a frame rotating at $\Delta = \omega_d - \omega_c$ gives

\beq
\frac{\dif}{\dif t} \hat{a}(t) = (+i\Delta - \kappa/2)\hat{a}(t) - i G \hat{x}(t)\hat{a}(t) + \sqrt{\kappa_{in}}s_{in}(t) + \sqrt{\kappa_0}\delta\hat{s}_{vac}(t) \; ,
\label{eq:dadt}
\eeq

\beq
\frac{\dif}{\dif t} \hat{x}(t) = \frac{\hat{p}(t)}{m_{eff}} \; ,
\label{eq:dxdt}
\eeq

\beq
\frac{\dif}{\dif t} \hat{p}(t) = -m_{eff} \Omega^2_m \hat{x}(t) - \hbar G \hat{a}^\dagger(t)\hat{a}(t) - \Gamma_m \hat{p}(t) + \delta \hat{F}_{th}(t) \; ,
\label{eq:dpdt}
\eeq
where $\delta\hat{s}_{vac}(t)$ and $\delta \hat{F}_{th}(t)$ are the quantum and thermal noise respectively, $\kappa$ is the total cavity decay rate, $\kappa_0$ the internal cavity decay rate, and $\Gamma_m$ the mechanical decay rate. The static cavity field and mechanical displacement are then

\beq
\bar{a} = \frac{\sqrt{\kappa_{in}}}{\kappa/2 - i(\Delta - G \bar{x})}\bar{s}_{in}=\frac{\sqrt{\kappa_{in}}}{\kappa/2 -i\bar{\Delta}}\bar{s}_{in} \; ,
\eeq

\beq
\bar{x} = \frac{-\hbar G \abs{\bar{a}}^2}{m_{eff}\Omega_m^2} \; .
\eeq
Here, $\bar{\Delta} = \Delta - G \bar{x}$ is the `corrected detuning', accounting for the average radiation pressure force. For a single probe beam and weak or detuned control fields, $\bar{a}$ can be assumed real \cite{weis2010optomechanically}; however, for multiple probe beams $\bar{a}$ must be considered complex. Linearizing Eqs.~\ref{eq:dadt}, \ref{eq:dxdt} and \ref{eq:dpdt} for small perturbations using the ansatz $\hat{a}(t) = \bar{a} + \delta\hat{a}(t)$ and $\hat{x}(t) = \bar{x} + \delta\hat{x}(t)$, while retaining only first-order terms in the small quantities $\delta \bar{a}$, $\delta \bar{a}^\dagger$ and $\delta \bar{x}$ gives

\beq
\frac{\dif}{\dif t} \delta\hat{a}(t) = (+i\bar{\Delta} - \kappa/2)\delta\hat{a}(t) - i G \bar{a} \delta\hat{x}(t) + \sqrt{\kappa_{in}}\delta s_{in}(t) + \sqrt{\kappa_0}\delta\hat{s}_{vac}(t) \; ,
\label{eq:ddadt}
\eeq

\beq
\frac{\dif^2}{\dif t^2} \delta\hat{x}(t) + \Gamma_m \frac{\dif}{\dif t} \delta\hat{x}(t) + \Omega_m^2 \delta\hat{x}(t) = - \frac{\hbar G}{m_{eff}} (\bar{a}^\dagger \delta\hat{a}(t) + \bar{a} \delta\hat{a}^\dagger(t)) + \delta \hat{F}_{th}(t) \; .
\label{eq:ddxdt}
\eeq
Now considering two probe tones produced by amplitude modulation of the form $\cos(\Omega_p t)$, where $\pm \Omega_p = \omega_p^\pm - \omega_d$ is the separation in frequency between the pump and probe tones $\omega_p^\pm$, the perturbation to the incident field can be written as

\beq
\delta s_{in} = \frac{s_p}{2} \left(\me^{-i\Omega_p t} + \me^{+i\Omega_p t} \right) \; .
\label{eq:deltasin}
\eeq

Considering the microwave drives as classical coherent fields, expectation values can be used in place of operators $z(t) \equiv \expval {\hat{z}(t)}$, allowing both noises, which average to zero, to be dropped. For the form of $\delta s_{in}$ in Eq.~\ref{eq:deltasin}, a general solution can be found using the ansatz

\beq
\delta a(t) = A^+ \me^{-i\Omega_p t} + A^- \me^{+i\Omega_p t} \; ,
\label{eq:ansatza}
\eeq

\beq
\delta a^*(t) = (A^-)^* \me^{-i\Omega_p t} + (A^+)^* \me^{+i\Omega_p t} \; ,
\label{eq:ansatzastar}
\eeq

\beq
\delta x(t) = X \me^{-i\Omega_p t} + X^* \me^{+i\Omega_p t} \; ,
\label{eq:ansatzx}
\eeq
where Hermitian $\delta x (t)$ has been used. Note $A^+$ is used for $\exp(-i\Omega_p t)$ to refer to the upper sideband with $\exp(-i(\omega_d + \Omega_p)t)$. Substituting this ansatz into Eq. \ref{eq:ddadt} and Eq. \ref{eq:ddxdt} and sorting by frequency will yield six equations; for a single probe tone, only the three terms at the probe frequency are required \cite{weis2010optomechanically}, while for two probe tones, all six are necessary. The two $\delta a(t)$ terms are

\beq
(-i(\bar{\Delta} + \Omega_p) + \kappa/2)A^+ = -i G \bar{a} X + \sqrt{\kappa_{in}}s_p/2 \; ,
\label{eq:da-om}
\eeq

\beq
(-i(\bar{\Delta} - \Omega_p) + \kappa/2)A^- = -i G \bar{a} X^* + \sqrt{\kappa_{in}}s_p/2 \; ,
\label{eq:da+om} \; 
\eeq
while the $\delta a^*(t)$ terms are

\beq
(+i(\bar{\Delta} - \Omega_p) + \kappa/2)(A^-)^* = + i G \bar{a}^* X + \sqrt{\kappa_{in}}s_p/2 \; ,
\label{eq:dastar-om}
\eeq

\beq
(+i(\bar{\Delta} + \Omega_p) + \kappa/2)(A^+)^* = + i G \bar{a}^* X^* + \sqrt{\kappa_{in}}s_p/2 \; ,
\label{eq:dastar+om}
\eeq
and finally the $\delta x (t)$ terms are

\beq
m_{eff}(\Omega^2_m - \Omega_p^2 - i \Gamma_m \Omega_p)X = -\hbar G (\bar{a}^* A^+ + \bar{a}(A^-)^*) \; ,
\label{eq:dx-om}
\eeq

\beq
m_{eff}(\Omega^2_m - \Omega_p^2 + i \Gamma_m \Omega_p)X^* = -\hbar G (\bar{a}^* A^- + \bar{a}(A^+)^*) \; .
\label{eq:dx+om}
\eeq
These six equations can be used to solve for $A^+$ and $A^-$, but first, we define the cavity and mechanical susceptibilities as

\beq
\chipm = \frac{1}{-i(\bar{\Delta} \pm \Omega_p) + \kappa/2} \; ,
\eeq

\beq
\chi_m = \frac{1}{m_{eff}}\frac{1}{\Omega_m^2 - \Omega_p^2 - i\Omega_p\Gamma_m} \; .
\eeq
Solving Eqs. \ref{eq:da-om}, \ref{eq:dastar-om} and \ref{eq:dx-om} for $A^+$, and substituting in the susceptibilities, gives

\beq
\sqrt{\kappa_{in}}s_p/2 - \frac{A^+}{\chi_+} = -i\hbar G^2 \abs{\bar{a}}^2 \chi_m  \left[A^+ + \chi_-^* \left(1+\frac{\bar{a}^2}{ \abs{\bar{a}}^2}\right)\sqrt{\kappa_{in}}s_p/2 \; - \; \frac{\chi_-^*}{\chi_+}A^+ \right] \; ,
\label{eq:Aplusmessy}
\eeq
which allows for the substitution near the mechanical resonance ($\Omega_m \approx \Omega_p$) of

\beq
\begin{split}
\hbar G^2 \abs{\bar{a}}^2 \chi_m & = \frac{\hbar G^2 \abs{\bar{a}}^2}{m_{eff}} \frac{1}{\Omega_m^2 - \Omega_p^2 - i\Omega_p\Gamma_m}\\
& \approx \frac{\hbar G^2 \abs{\bar{a}}^2}{2 \Omega_m m_{eff}} \frac{1}{\Omega_m - \Omega_p - i \Gamma/2}\\
& = \frac{\abs{g}^2}{-\delta - i\Gamma_m/2} \;,
\label{eq:Gsubg}
\end{split}
\eeq
using $\delta = \Omega_p - \Omega_m$ and $g = \bar{a}g_0$, the photon-enhanced electromechanical coupling in the linearized regime, where $g_0 = x_{zpf}G$ is the vacuum electromechanical coupling, equivalent to the shift in cavity resonance due to the zero point motion of the mechanical resonator. The standard definition of zero point motion $x_{zpf} = \sqrt{\hbar/2\Omega_m m_{eff}}$ is used here. Substituting Eq.~\ref{eq:Gsubg} into Eq.~\ref{eq:Aplusmessy} and rearranging gives

\beq
A^+  = \chi_+ \left( 1 + \frac{i\abs{g}^2 \left(\frac{\bar{a}^2}{\abs{\bar{a}}^2} \chi_-^* + \chi_+\right)}{
-\delta - i\Gamma_m/2 \; + \; i\abs{g}^2(\chi_-^* - \chi_+)}\right) \sqrt{\kappa_{in}}s_p/2 \; .
\label{eq:Aplusold}
\eeq
The same method can be used to solve for $A^-$ using Eqs. \ref{eq:da+om}, \ref{eq:dastar+om} and \ref{eq:dx+om}, resulting in

\beq
A^- = \chi_- \left( 1 + \frac{i\abs{g}^2 \left(\frac{\bar{a}^2}{\abs{\bar{a}}^2} \chi_+^* + \chi_-\right)}{
-\delta + i\Gamma_m/2 \; + \; i\abs{g}^2(\chi_+^* - \chi_-)}\right) \sqrt{\kappa_{in}}s_p/2 \; .
\label{eq:Aminusold}
\eeq
These forms are similar to other OMIT/A experiments \cite{weis2010optomechanically,zhou2013slowing}. To obtain forms equivalent to Kashkanova \etal~\cite{kashkanova2017optomechanics} define the electromechanical self-energy as

\beq
i\Sigma(\Omega_p) = \abs{g}^2(\chi_+ - \chi_-^*) \; .
\eeq
This also defines the electromechanical spring effect $\Delta \Omega_{m(ele)} = \ree{\Sigma(\Omega_p)}$, and the electromechanical damping $\Gamma_{m(ele)} = \imm{\Sigma(\Omega_p)}$. Using the self-energy and $g = \bar{a}/ \bar{a}^* g^*$, Eq.~\ref{eq:Aplusold} and Eq.~\ref{eq:Aminusold} can be rewritten as

\beq
A^+ = \chi_+ \sqrt{\kappa_{in}}s_p/2 \; - \; \frac{g\chi_+(g\chi_-^* +g^*\chi_+)}{\Gamma_m/2 - i\delta + i\Sigma(\Omega_p)}\sqrt{\kappa_{in}}s_p/2 \; ,
\label{eq:Aplus}
\eeq

\beq
A^- = \chi_- \sqrt{\kappa_{in}}s_p/2 \; + \; \frac{g\chi_-(g\chi_+^* +g^*\chi_-)}{\Gamma_m/2 + i\delta - i\Sigma(-\Omega_p)}\sqrt{\kappa_{in}}s_p/2 \; ,
\label{eq:Aminus}
\eeq
which are $A^\pm$ in equation (6) of the main paper, effectively the cavity field amplitudes at $\omega_p^\pm$ and equivalent to $a_+[\Omega]$ and $a_-[\Omega]$ in Ref.~\cite{kashkanova2017optomechanics}.

\subsection{Canceling The Probe Tones}

This subsection will take the more general expressions for microwave field amplitude $A^+$ and $A^-$, and develop the specific technique of three-tone measurement used in this work. Using the standard input-output relationship for a single port cavity $\hat{a}_{out} = \hat{a}_{in} - \sqrt{\kappa_{in}} \hat{a}$ \cite{aspelmeyer2014cavity}, the choice of $\delta s_{in}$ in Eq.~\ref{eq:deltasin}, and using $s_{in}(t) = \left(\bar{s}_{in} + \delta s (t)\right)\me^{-i\omega_d t}$, the signal reflected from the cavity $s_{out}(t)$ is

\beq
\begin{split}
s_{out}(t) = & \; (\bar{s}_{in} -  \sqrt{\kappa_{in}}\bar{a})\me^{-i(\omega_d t + \phi)} \\
& + ( s_p/2 -  \sqrt{\kappa_{in}} A^+ ) \me^{-i([\omega_d + \Omega_p] t + \phi + \psi)}  \\
& + ( s_p/2 -  \sqrt{\kappa_{in}} A^- ) \me^{-i([\omega_d - \Omega_p] t + \phi - \psi)}  \; .
\end{split}
\eeq
Demodulating this signal with frequency $\omega_d$ and phase $\phi_{ref}$, using a frequency mixer that effectively multiplies by $\cos(\omega_d t + \phi_{ref})$, then filtering out the $2\omega_d$ and DC terms gives

\beq
s_{demod}(t) = \frac{K_T}{2} \left[ ( s_p - \sqrt{\kappa_{in}} A^+ )\me^{-i(\Omega_p t + \psi)} + ( s_p - \sqrt{\kappa_{in}} A^- ) \me^{+i(\Omega_p t + \psi)} \right] \me^{i(\phi_{ref} - \phi)} \; ,
\label{eq:sdemod}
\eeq
where $k_p$ is the transfer function of the circuit after the cavity, including the mixer. The additional 10 kHz amplitude modulation (far from $\Omega_m$) sets $\phi_{ref} - \phi = \pi/2$. A double-balanced mixer is used to minimize loss, rather than an IQ mixer as the probe tones in the opposing quadrature would drown out any coherent signal. As a double-balanced mixer is used only the real part of the signal is kept; therefore, expanding the exponentials in $s_{demod}(t)$ and keeping only the real terms gives

\beq 
\begin{split}
s_{demod}^\mathbb{R} (t) /K_T & = - \ree{i(A^+  +  A^-)} \frac{\sqrt{\kappa_{in}}}{2} \cos(\Omega_p t + \psi) +  \ree{A^- - A^+} \frac{\sqrt{\kappa_{in}}}{2} \sin(\Omega_p t + \psi) \\
& = \imm{A^+  +  A^-} \frac{\sqrt{\kappa_{in}}}{2} \cos(\Omega_p t + \psi) +  \ree{A^- - A^+} \frac{\sqrt{\kappa_{in}}}{2} \sin(\Omega_p t + \psi) \; ,
\label{eq:sdemodR}
\end{split}
\eeq
which is the signal the lock-in amplifier then measures.
\subsection{Unresolved-Sideband Limit}
The key regime of interest here is deep into sideband-unresolved ($\kappa \gg \Omega_m$), and while the technique is also effective for stronger couplings this work focuses on weak an electromechanical coupling ($\Delta\Omega_{m(ele)} \ll \Gamma_m$), where standard EMIT/A cannot provide sufficient sensitivity. A sideband-resolved two-probe measurement scheme can be found in Ref.~\cite{buters2017straightforward}, where the cavity filters out the pump and lower probe tone. To simplify the measured signal for the sideband-unresolved regime, first approximate the cavity susceptibilities (using $\Omega_p^2,\Omega_p\Delta \ll \kappa^2$) to

\beq
\chipm  \approx \frac{\kappa/2 + i (\Delta \pm \Omega_p)}{\kappa^2/4 + \Delta^2} \; ,
\label{eq:chiapprox}
\eeq
where $\bar{\Delta} \rightarrow \Delta$ has been used, as the coupling is expected to be weak. Using the approximation in Eq.~\ref{eq:chiapprox} the terms of interest $\imm{A^+  +  A^-}$ and $\ree{A^- - A^+}$ can be written:

\beq
\imm{A^+  +  A^-} = \frac{\Delta\sqrt{\kappa_{in}}s_p}{\kappa^2/4 + \Delta^2} + \frac{\sqrt{\kappa_{in}}s_p}{2} \; \imm{\frac{g\chi_-(g\chi_+^* +g^*\chi_-)}{\Gamma_m/2 + i\delta - i\Sigma(-\Omega_p)} - \frac{g\chi_+(g\chi_-^* +g^*\chi_+)}{\Gamma_m/2 - i\delta + i\Sigma(\Omega_p)}} \; ,
\label{eq:immpart}
\eeq

\beq
\ree{A^- -  A^+} = \frac{\sqrt{\kappa_{in}}s_p}{2} \; \ree{\frac{g\chi_-(g\chi_+^* +g^*\chi_-)}{\Gamma_m/2 + i\delta - i\Sigma(-\Omega_p)} + \frac{g\chi_+(g\chi_-^* +g^*\chi_+)}{\Gamma_m/2 - i\delta + i\Sigma(\Omega_p)}} \; .
\label{eq:reepart}
\eeq
To expand further, use the full form of the photon-enhanced coupling

\beq
\begin{split}
g & = \frac{g_0 \sqrt{\kappa_{in}} \bar{s}_{in}}{\kappa/2 - i\Delta} \\
& = g_0 \sqrt{\kappa_{in}} \bar{s}_{in}\frac{\kappa/2 + i\Delta}{\kappa^2/4 + \Delta^2} \; ,
\end{split}
\eeq
along with Eq.~\ref{eq:chiapprox} and the expected small self-energy $\Sigma (\Omega_p) \ll \Gamma_m$, to expand the factors within $\imm{...}$ and $\ree{..}$ in Eq.~\ref{eq:immpart} and Eq.~\ref{eq:reepart}, giving

\beq
\frac{g(g\chi_- \chi_+^* + g^*\chi_-^2)}{\Gamma_m/2 + i\delta} = g_0^2 \kappa_{in} \bar{s}_{in}^2 (\kappa/2 + i\Delta) 
\frac{\kappa^3/4 + \kappa\Delta^2 + \kappa\Delta\Omega_p + i(2 \Delta^3 + \kappa^2 \Delta/2 -\kappa^2\Omega_p - 2\Delta^2\Omega_p)}{\left(\kappa^2/4 + \Delta^2\right)^4(\Gamma/2 + i \delta)} \; ,
\eeq

\beq
\frac{g(g \chi_+ \chi_-^* + g^* \chi_+^2)}{\Gamma_m/2 - i\delta}  = g_0^2 \kappa_{in} \bar{s}_{in}^2 (\kappa/2 + i\Delta) 
\frac{\kappa^3/4 + \kappa\Delta^2 - \kappa\Delta\Omega_p + i(2\Delta^3 + \kappa^2 \Delta/2 + \kappa^2\Omega_p + 2\Delta^2\Omega_p)}{\left(\kappa^2/4 + \Delta^2\right)^4(\Gamma/2 - i \delta)} \; ,
\eeq
where factors of $\Omega_p^2 \; (\ll \kappa^2)$ have been discarded. Substituting these equations into Eq.~\ref{eq:immpart} and Eq.~\ref{eq:reepart}, then taking the imaginary and real parts respectively gives

\beq
\imm{A^+  +  A^-} = \frac{\Delta\sqrt{\kappa_{in}}s_p}{\kappa^2/4 + \Delta^2} - \frac{\sqrt{\kappa_{in}}s_p}{2} g_0^2 \kappa_{in} \bar{s}_{in}^2 
\frac{(\kappa^3\Omega_p)\Gamma_m/2 + (\kappa^4/4 - 4\Delta^4)\delta}{\left(\kappa^2/4 + \Delta^2\right)^4(\Gamma^2/4 + \delta^2)} \; ,
\eeq

\beq
\ree{A^- - A^+} = \frac{\sqrt{\kappa_{in}}s_p}{2} g_0^2 \kappa_{in} \bar{s}_{in}^2
\frac{(\kappa^4/4 - 4\Delta^4)\Gamma_m/2 - (\kappa^3\Omega_p)\delta}{\left(\kappa^2/4 + \Delta^2\right)^4(\Gamma^2/4 + \delta^2)} \; .
\eeq
Ignoring the first term in $\imm{A^+  +  A^-}$ as a constant background which can be subtracted, and using $\Omega_p \ll \kappa$ to remove all $\Omega_p$ terms, $s_{demod}^\mathbb{R} (t)$ can be written as

\beq
s_{demod}^\mathbb{R} (t) = K_T g_0^2 \kappa_{in}^2 s_p \bar{s}_{in}^2 
\frac{(\kappa^2/4 - \Delta^2)}{\left(\kappa^2/4 + \Delta^2\right)^3(\Gamma^2/4 + \delta^2)} [\sin(\Omega_p t + \psi)\Gamma_m/2 - \cos(\Omega_p t + \psi) \delta] \; ,
\eeq
which is the signal reaching the measurement lock-in amplifier. The lock-in takes a signal of form $V_s(t) = A_s \cos(\Omega_s t + \theta)$ and returns

\beq
X+iY = \frac{A_s}{\sqrt{2}}F(\Omega_s-\Omega_{ref})\exp{i[(\Omega_s-\Omega_{ref}t + \theta]} \; .
\eeq 
So measuring $s_{demod}^\mathbb{R} (t)$ at frequency $\Omega_p$ will return a signal

\beq
X + i Y = \frac{-iK_T g_0^2 \kappa_{in}^2 s_p}{\sqrt{2}(\Gamma/2 + i\delta)} 
\frac{(\kappa^2/4 - \Delta^2)}{\left(\kappa^2/4 + \Delta^2\right)^3} \frac{P_{in}}{\hbar \omega_d} \; ,
\label{eq:xplusiysi}
\eeq
where $\bar{s}_{in}^2 = P_{in}/\hbar\omega_d$. This finally recovers equation (10) in the main text, though $K_T$ is still unknown which prevents calibration of $g_0$ from a single measurement. The phase $\me^{i\psi}$ has been dropped as it will be small for a low mechanical frequency; in practice, adjustment of the lock-in phase can always set $\psi = 0$.

\subsection{Calibration Of Vacuum Electromechanical Coupling}

To calibrate the three-tone EMIT/A response a measurement of the probe tones interfering constructively is used. This allows for a correction of equation (10) in the main text for $K_T$ and $s_p$, enabling calculation of $g_0$. The calibration is achieved by setting $\phi_{ref}-\phi = 0$ in Eq.~\ref{eq:sdemod} for $s_{demod}$, giving

\beq
s_{demod}^{\mathbb{R},cal}(t)/K_T = \ree{s_p - \sqrt{\kappa_{in}}(A^+  +  A^-)}\frac{1}{2} \cos(\Omega_p t + \psi) + \imm{(A^- - A^+)} \frac{\sqrt{\kappa_{in}}}{2} \sin(\Omega_p t + \psi) \; .
\eeq
The coherent contribution to the signal from the electromechanical interaction can be ignored here, either because it is small compared to the probe tones (for weak coupling) or because the calibration can be taken detuned from the mechanical resonance. The sine term is also small compared to the cosine as $\Omega_m \ll \kappa$. Therefore the signal measured at the lock-in is just the combined probe tones reflected from the cavity (including the cavity susceptibility), which can be written as

\beq
X+iY =\frac{s_p K_T}{2\sqrt{2}}\left(1- \frac{\kappa_{in}\kappa/2}{\kappa^2/4 + \Delta^2}\right)\me^{i\psi} \; .
\eeq

The reference value for calibration can be achieved by correcting for the bracketed term, using fits of the microwave resonance and taking the magnitude, giving

\beq
R_{cal} = \frac{\abs{X+iY}}{1- \frac{\kappa_{in}\kappa/2}{\kappa^2/4 + \Delta^2}} = \frac{s_p K_T}{2\sqrt{2}} \; .
\eeq
Normalizing $X+iY$ in Eq.~\ref{eq:xplusiysi} by $R_{cal}$, taking the value at $\Gamma_m$ and rearranging, allows calculation of $g_0$ according to

\begin{equation}
    g_0 = \frac{\sqrt{\hbar\omega\ts d}}{2 \kappa\ts{in} \sqrt{P\ts{in}}} \sqrt{\frac{\Gamma_m(\kappa^2/4 + \Delta^2)^3}{(\kappa^2/4 - \Delta^2)} \frac{R_{\delta =0}}{R_{cal}}} \; ,
\end{equation}
which is equation (11) in the main text. The magnitude of the coherent signal at $\delta = 0$ is $R_{\delta =0}$, either from the maximum of the coherent spectrum when sweeping $\Omega_p$ or calculated via fitting of the same spectrum. Backgrounds should be subtracted from calculations of $R_{\delta =0}$, as these normally depend on contributions from imperfect phase-locking. To calculate $R_{cal}$, the phase $\phi_{ref}$ is swept and the maximum of the magnitude is taken.

\section{Homodyne Measurement Circuit}

\begin{figure}[h!]
  \centering
  \includegraphics[width=0.8\textwidth]{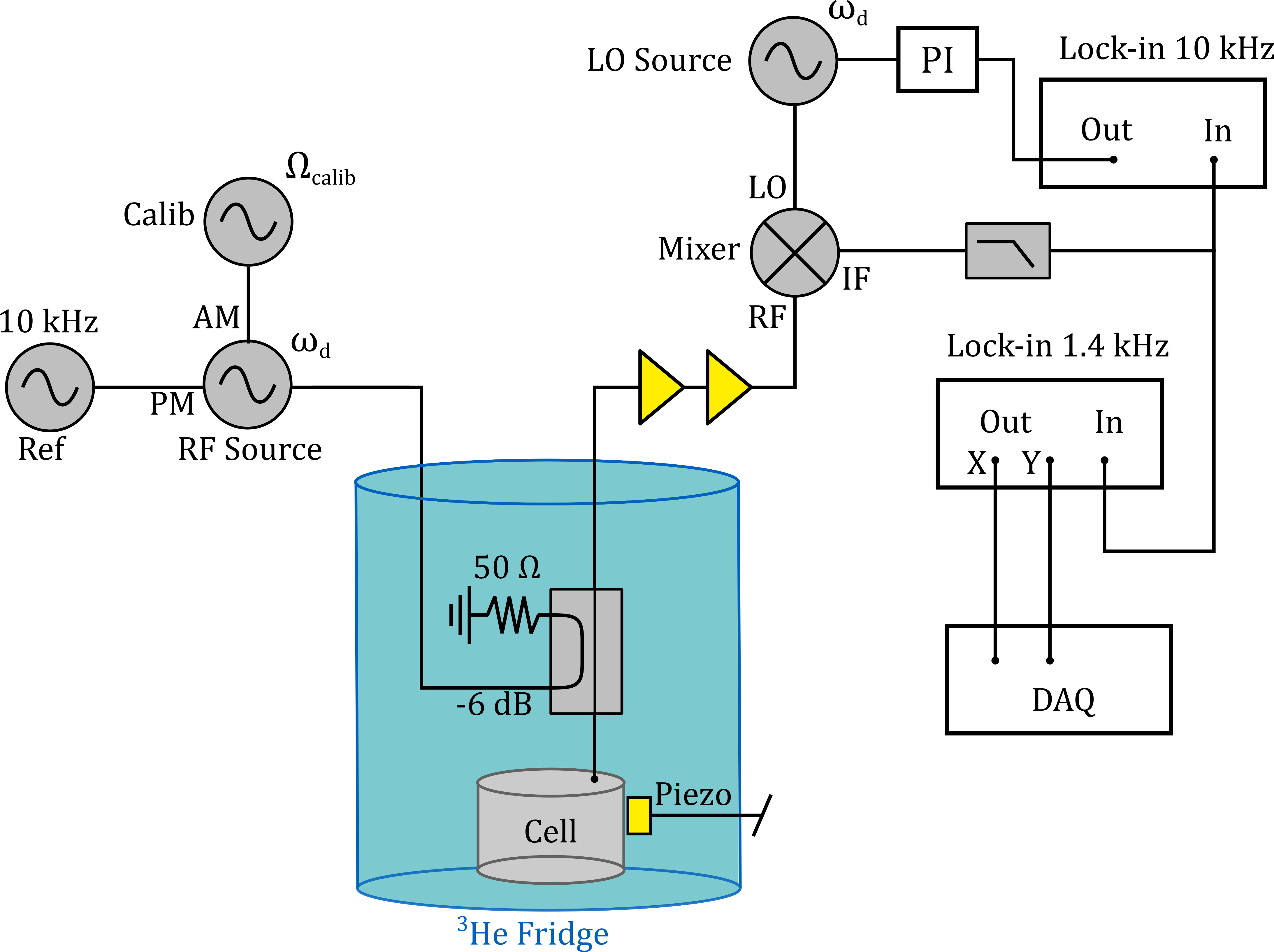}
  \caption{Amplitude-sensitive calibrated homodyne measurement circuit. The RF Source microwave signal, at frequency $\omega_d$, is phase modulated with the 10 kHz reference signal and amplitude modulated with the calibration tone at $\Omega_{calib}$. The RF signal is sent to the experimental `Cell' via a directional coupler. The reflected signal passes back through the directional coupler, is amplified with two microwave amplifiers, and is mixed with the LO source signal at frequency $\omega_d$. The mixed signal is low-pass filtered to remove the $2\omega_d$ components and is then measured on two lock-in amplifiers. The lower lock-in on the diagram is set to 1.4 kHz, near the mechanical frequency $\Omega_m$, and acts as a filter plus amplifier. The $X$ and $Y$ quadrature outputs of this lock-in are measured on a DAQ card, giving a time domain noise measurement of frequency components close to $\Omega_m$. The upper lock-in on the diagram measures the 10 kHz reference signal, which is again used to lock the relative phase of mixing via the `LO Source'. A piezo is present on the outside of the cell to allow for mechanical driving, either with swept frequency to find mechanical modes, or with white noise to raise the effective mode temperature.}
  \label{fig:homocircuit}
\end{figure}

Figure~\ref{fig:homocircuit} shows the microwave measurement setup for amplitude-sensitive calibrated homodyne measurement. The method here broadly follows that of Ref.~\cite{gorodetksy2010determination}, adapted for microwave circuits in Ref.~\cite{spence2022superfluid,kumar2023novel}. This method allows readout of the Helmholtz resonator's mechanical noise spectral density, as this noise imparts a time dependence on the microwave resonance via electromechanical coupling. The time-dependent cavity resonance then imparts noise on a pump signal, measured in the time domain by down-mixing with an LO of the same frequency, and Fourier transformed to give the mechanical noise spectral density. The figure shows an amplitude-sensitive scheme, where $\omega_d$ is placed at the maximum gradient of the microwave resonance, the 10 kHz reference is applied via phase modulation, and the calibration tone via amplitude modulation. Phase-sensitive measurement was also used to measure the mechanics, where $\omega_d$ is placed at the maximum amplitude of the microwave resonance, the 10 kHz reference is applied via amplitude modulation, and the calibration tone via phase modulation.

Working backward from $g_0$, and using the calibrated homodyne technique \cite{gorodetksy2010determination,spence2022superfluid}, an estimate for effective Helmholtz mode temperature is calculated giving $T\ts{eff} \approx 80,000$ K, which demonstrates the strong incoherent driving from the cryocooler. While initially, this figure seems high, a silicon nitride membrane driven with white noise via a piezo at 0.001 mV$^2$/Hz had an effective temperature of 23,000 K \cite{kumar2023novel}; meanwhile, driving the Helmholtz resonator at over 2.5 mV$^2$/Hz in this work could not overcome the cryocooler.

